\documentclass[prl,reprint,showpacs,showkeys,showpacs]{revtex4-1}
\usepackage{amsmath,amssymb,amsthm,times,graphics,graphicx,bm,subfigure}
\usepackage[colorlinks,linkcolor=red]{hyperref}

\newcommand{\be}{\begin{equation}}
\newcommand{\ee}{\end{equation}}
\newcommand{\ben}{\begin{eqnarray}}
\newcommand{\een}{\end{eqnarray}}
\newcommand{\bes}{\begin{subequations}}
\newcommand{\ees}{\end{subequations}}

\def\ket#1{ | #1 \rangle}
\def\bra#1{{\langle #1 | }}
\def\tr{ {\rm{Tr }}}

\newcommand{\proj}[1]{\mbox{$|#1\rangle \!\langle #1 |$}}

\newcommand{\arxiv}[2][arxiv:]{\href{http://arxiv.org/abs/#1#2}{#1#2}}

\newtheorem{theorem}{Theorem}
\newtheorem{lemma}{Lemma}

\begin{document}
\title{A Condition for the Nullity of Quantum Discord}
\author{Animesh Datta}
\email{animesh.datta@physics.ox.ac.uk}
\affiliation{Clarendon Laboratory, Department of Physics, University of Oxford, OX1 3PU, United Kingdom}

\date{\today}
\begin{abstract}
The positivity of quantum discord is shown to be equivalent to the strong subadditivity of von-Neumann entropy. This leads us to a necessary and sufficient condition characterizing the set of states with zero quantum discord. This also gives us a mathematical definition of pointer states, as they are the states with zero discord. Finally, we suggest that strong subadditivity of entropy might delineate the boundaries of the set of quantum correlations.
\end{abstract}



\maketitle


Identification, characterization and manipulation of nonclassical correlations is at the heart of quantum information science. Canonically, this has meant the study of quantum entanglement~\cite{hhhh09,pv07}. The role of entanglement is, however, far from clear in its most celebrated application, quantum computation. In particular, instances of mixed state quantum computation exist which outdo their best classical counterparts with limited amounts of entanglement~\cite{datta05a,datta07a}. This motivated the study of other measures of capturing nonclassical correlations, the foremost of which has been quantum discord~\cite{henderson01a,ollivier01a}. It has been studied in a variety of systems~\cite{datta08a,Dillenschneider08,datta09a,bwvu09} and circumstances~\cite{rmkss08,mcsv09,sl09,fwbac09}. This has subsequently led to the introduction of several related measures~\cite{Luo08a,wpm09}, inspired by different physical~\cite{bt10} and operational principles~\cite{mpsvw10}.


Quantum discord aims at capturing all quantum correlations in a quantum state, including entanglement~\cite{ollivier01a,henderson01a}. Quantum mutual information is generally taken to be the measure of total correlations, classical and quantum, in a quantum state. For two systems, $A$ and $B$, it is defined as $ I(A:B) = H(A) + H(B) -H(A,B),$ where $H(\cdot)$ stands for the von Neumann entropy, $H(\rho)\equiv -\tr(\rho\log\rho).$ In our paper, all logarithms are taken to base 2. For a classical probability distribution, Bayes' rule leads to an equivalent definition of the mutual information as $I(A:B) = H(A)-H(A|B),$ where the conditional entropy $H(A|B)$ is an average of the Shannon entropies of $A,$ conditioned on the alternatives of $B.$ It captures the ignorance in $A$ once the state of $B$ has been determined. For a quantum system, this depends on the measurements that are made on $B.$ For a POVM given by the set $\{\Pi_i\},$ the state of $A$ after the measurement corresponding to the outcome $i$ is given by
\be
\label{eq:postmeas}
\rho_{A|i} = \tr_B(\Pi_i\rho_{AB})/p_i,\;\;\;p_i=\tr_{A,B}(\Pi_i\rho_{AB}).
\ee
A quantum analogue of the conditional entropy can then be defined as $\tilde{H}_{\{\Pi_i\}}(A|B)\equiv\sum_ip_iH(\rho_{A|i}),$ and an alternative version of the quantum mutual information can now be defined as $\mathcal{J}_{\{\Pi_i\}}(A:B) = H(A)-\tilde{H}_{\{\Pi_i\}}(A|B).$ The above quantity depends on the chosen set of measurements $\{\Pi_i\}.$ To capture all the classical correlations present in $\rho_{AB},$ we maximize $\mathcal{J}_{\{\Pi_i\}}(A:B)$ over all $\{\Pi_i\},$ arriving at a measurement independent quantity $\mathcal{J}(A:B) = \max_{\{\Pi_i\}}(H(A)-\tilde{H}_{\{\Pi_i\}}(A|B)) \equiv H(A)-\tilde{H}(A|B),$ where $\tilde{H}(A|B)=\min_{\{\Pi_i\}}\tilde{H}_{\{\Pi_i\}}(A|B).$
Since the conditional entropy is concave over the set of POVMs, which is convex, the minimum is attained on the extreme points of the set of POVMs, which are rank 1~\cite{dattathesis}. Then, quantum discord is finally defined as
\ben
\label{discexp}
\mathcal{D}(A:B) &=& I(A:B)-\mathcal{J}(A:B) \\
                 &=& H(A)-H(A:B)+\min_{\{\Pi_i\}}\tilde{H}_{\{\Pi_i\}}(A|B),\nonumber
\een
where $\{\Pi_i\}$ are now, and henceforth in the paper, rank 1 POVMs. It is well known that the quantum discord is non-negative for all quantum states~\cite{ollivier01a,dattathesis}. In this paper, we present a new proof of this fact, based on the strong subadditivity of the von-Neumann entropy. More importantly, it allows us to derive the first necessary and sufficient condition for a state to have zero quantum discord. They are the ones that saturate strong subadditivity, and are infact, quite easy to work with, as we see later. This leads us to suggest an experimental realization of pointer states in superconducting cavity QED systems. Finally, we propose a simple certificate for distinguishing classical, quantum and more general no-signalling theories based on strong subadditivity.


\begin{theorem}
 \label{discordispositove} Quantum discord is always positive, \textit{\textit{i.e.}}, $\mathcal{D}(A,B)\geq 0.$
\end{theorem}

\noindent {\it Proof: } Consider the joint state $\rho_{AB}$ subject to one dimensional orthogonal measurements $\Pi_j=\proj{e_j}$ on $B$ (extended to arbitrary, at most $dim(B)^2$ dimensions, by the Neumark extension). The post-measurement state is given by Eq.~(\ref{eq:postmeas}). Suppose now that a system $C$ interacts with  $B$ so as to make the desired measurement ($U\ket{e_j}\otimes\ket{0}=\ket{e_j}\otimes\ket{f_j}$), leaving the state
 \be
 \label{extrho}
 \rho'_{ABC}=\sum_{j,k}\bra{e_j}\rho_{AB}\ket{e_k}\otimes\ket{e_j}\bra{e_k}\otimes\ket{f_j}\bra{f_k}.
 \ee
If the eigendecomposition of $ \rho_{AB} =\sum_l\lambda_l\proj{r_l} $, then
 \ben
 \rho'_{ABC}&=&\sum_{j,k,l}\lambda_l\langle \mathbb{I}_A,e_j\proj{r_l}\mathbb{I}_A,e_k\rangle \otimes\ket{e_j} \bra{e_k}\otimes\ket{f_j}\bra{f_k} \nonumber\\
            &=& \sum_{l}\lambda_l\proj{e_l,r_l,f_l}\nonumber
\een
whereby $ H(\rho'_{ABC})=H(\rho_{AB}).$ Also, from Eq.~(\ref{extrho}),
 \bes
\label{partials}
 \ben
 \rho'_{AB}&=&\sum_{j}p_j\rho_{A|j}\otimes\proj{e_j},  \\
 \rho'_{BC}&=&\sum_{j,k}\proj{e_j}\rho_B\proj{e_k}\otimes\ket{f_j}\bra{f_k}, \\
 \rho'_{B}&=&\sum_{j}p_j\proj{e_j},.
 \een
 \ees
whereby $H(\rho'_{AB})=H({\bm p})+\sum_j p_jH(A|j), H(\rho'_{BC})=H(\rho_B)$ and $H(\rho'_{B})=H({\bm p})$ respectively. These reduce the strong subadditivity of the von-Neumann entropy~\cite{lr73}
 \be
 \label{ssa}
 H(\rho'_{ABC}) + H(\rho'_{B}) \leq H(\rho'_{AB}) + H(\rho'_{BC}),
 \ee
to $ H(\rho_{AB}) + H({\bm p}) \leq H({\bm p})+\sum_j p_jH(A|j) + H(\rho_B),$ whereby
 \ben
 \tilde H_{\{\Pi_j\}}(A|B) &\equiv& \sum_j p_jH(A|j) \nonumber \\
                            &\geq& H(\rho_{AB})- H(\rho_B) \equiv H(A|B).
 \een
This, being true for all measurements, also holds for the minimum, proving the theorem.
 \qed

This theorem also applies to infinite dimensional systems, since the von-Neumann entropy is strongly subadditive for such systems as well~\cite{lr73}, albeit the summations in Eqs.~(\ref{extrho}) and the following could be infinite. Having proven that the quantum discord is always nonnegative, it is evident that the condition for zero discord can be reduced to that of the equality in strong subadditivity in Eq.~(\ref{ssa}). To that end, we will employ a result stated by Ruskai~\cite{r02}, and proven by Hayden \textit{et al}.~\cite{hjpw04}, which we quote below for completeness.

\begin{lemma}[\cite{hjpw04}]
 \label{ssaeq}
A state $\rho'_{ABC}$ on $\mathcal{H}_A\otimes\mathcal{H}_B\otimes \mathcal{H}_C$ satisfies strong subadditivity (Eq.~(\ref{ssa})) with equality if and only if there is a decomposition of system $B$ as
 $$
 \mathcal{H}_B = \bigoplus_j \mathcal{H}_{B^L_j}\otimes \mathcal{H}_{B^R_j}
 $$
into a direct sum of tensor products such that
 $$
\rho'_{ABC} = \bigoplus_j q_j \rho_{AB^L_j}\otimes \rho_{B^R_j C}
  $$
with states $\rho_{AB^L_j}$ on $\mathcal{H}_A\otimes\mathcal{H}_{B^L_j}$ and $\rho_{AB^R_j}$ on $\mathcal{H}_{B^R_j}\otimes\mathcal{H}_C$, and a probability distribution $\{q_j\}.$
\end{lemma}

\begin{theorem}
\label{zerodiscord} $\mathcal{D}(A,B)= 0$ if and only if the state $\rho_{AB}$ is joint density operator is block-diagonal in the marginal eigenbasis of $B$, that is
 $$
 \rho_{AB} = \sum_{j} P_j \rho_{AB} P_j
 $$
where $\rho_{AB}=\sum_j \tau_j P_j$, with $\{\tau_j\}$ a probability distribution~\cite{ollivier01a}.
\end{theorem}

\noindent{\it Proof:} For any measurement on $B$ executed via $C$, the state has the form
 \be
\rho'_{ABC} = \sum_{\alpha}q_{\alpha} \rho_{A|\alpha}\otimes
\rho^{\alpha}_{BC},
 \ee
where $\rho^{\alpha}_{BC} =\Pi_{\alpha}\rho^{\alpha}_{BC}\Pi_{\alpha},$ with $\Pi_{\alpha}$ being projectors of the form $ \Pi_{\alpha} = \sum_{j}\proj{E_{\alpha j}}\otimes \proj{F_{\alpha j}}.$ In our case, the state $\rho'_{ABC}$ is invariant under the exchange of $B$ and $C$ relative to the measurement basis, here denoted by $\ket{E_{\alpha j}}$ and $\ket{F_{\alpha j}}$. Additionally, following Lemma.~(\ref{ssaeq}), for any measurement that saturates Eq. (\ref{ssa}), there exists a decomposition of the Hilbert space of $B$ that can be written as $ \mathbb{I}_B = \sum_{\alpha}\Pi_{\alpha} =\sum_{\alpha}\Pi_{\alpha L}\otimes \Pi_{\alpha R},$ and $\Pi_{\alpha}\Pi_{\beta}=\delta_{\alpha\beta}\Pi_{\alpha}.$
Thus, $ \rho^{\alpha}_{BC} = \sum_{j,k}\rho^{\alpha}_{jk}\ket{E_{\alpha j}}\bra{E_{\alpha k}}\otimes \ket{F_{\alpha j}}\bra{F_{\alpha k}}$ and $ \rho'_{AB} = \sum_{\alpha}q_{\alpha} \rho_{A|\alpha}\otimes \rho^{\alpha}_{B} = \sum_{\alpha,j} \rho^{\alpha}_{jj}q_{\alpha}\rho_{A|\alpha}\otimes \ket{E_{\alpha
j}}\bra{E_{\alpha j}}.$ Undoing the measurement, $U\ket{e_j}\otimes\ket{0}=\ket{e_j}\otimes\ket{f_j}$, gives
 $$
\rho_{AB}=\bra{0_C}U^{\dag}\rho'_{ABC}U\ket{0_C}=\sum_{\alpha}q_{\alpha}\rho_{A|\alpha}\otimes \rho^{\alpha}_{B}.
 $$
Diagonalizing $\rho^{\alpha}_B=\sum_k \lambda^{\alpha}_k\proj{\lambda^{\alpha}_k}$, we get $ \rho_{AB}=\sum_{\alpha,j}\lambda^{\alpha}_j
q_{\alpha}\rho_{A|\alpha}\proj{\lambda^{\alpha}_j}.$
Relabelling, we have that the discord is zero if and only if
 \be
\rho_{AB}=\sum_{j} p_j \rho_{A|j}\otimes \proj{\lambda_j}
 \ee
in the basis that diagonalizes $\rho_B.$ The $\alpha$ subspaces take into account that if the states $\rho_{A|j}$ are the same for
different $j$, then we can attain zero discord by using any measurement in the subspace spanned by those values of $j$.  Diagonalising $\rho_{A|j} = \sum_k \mu_{jk}\proj{\mu_{jk}}$, we get that a state has zero discord if and only if
 \be
\label{treebasis}
\rho_{AB} = \sum_{jk} p_j\mu_{jk}\proj{\mu_{jk},\lambda_j}.
 \ee
Thus the eigenbasis of $\rho_{AB}$ has a tree product structure $\ket{\mu_{jk}}\otimes\ket{\lambda_j}$.
\qed


The condition for equality in strong subadditivity is that the state $\rho'_{ABC}$ be what is known as a short Markov chain, that is, $C$ is conditionally independent from $A$, given $B.$ Since this also means that it is possible to extract information from $B$ using non-demolition measurements~\cite{hjpw04}, we have arrived at a formal mathematical characterization of pointer states, which formed the background for the introduction of quantum discord in the first place~\cite{ollivier01a}. These states form the boundary between quantum and classical correlations, particularly, as quantum discord quantifies the amount of information that cannot be extracted without joint measurements. Interestingly, the set of pointer states is of measure zero, as can be concluded from the results about the measure of the set of zero discord states~\cite{facca09}. Note that the definition of zero discord states presented above is also valid for infinite dimensional systems, provided $H(\rho'_{AB}), H(\rho'_B),H(\rho'_{BC})$ are finite~\cite{jp06,petz06}. It is however an interesting question whether the set of zero discord states or pointer states in continuous variable systems is of measure zero~\cite{ad10}.

\textit{Discussions--}
Since the states with vanishing discord were identified with pointer states, our results also allow for their mathematical and constructive definition. Curiously, they form a set of measure zero within the space of density matrices in finite dimensions. Finally, recent efforts of identifying information-theoretic criterion to differentiate quantum theory from more general no-signalling theories have led to the notion of information causality~\cite{ppkswz09}, which is respected by quantum and classical theories, but not by more general theories. These violations have now been traced to a violation of the strong subadditivity of entropy~\cite{sw10,bbclssww10}. This leads us to propose that strong subadditivity of entropy demarcates the edges of quantum correlations, with the quantity
 $$
 H(AB) + H(BC) - H(ABC) - H(B)
 $$
being positive for quantum states, zero for classical states, and negative for states in more general probabilistic theories. We hope that the connections unraveled in this paper, between pointer states, strong subadditivity, quantumness and discord lead to better understandings of the quantum nature of the universe, at a conceptual as well as a practical level.

The author acknowledges numerous discussions with C. M. Caves, A. Shaji, M. Lang, G. Adesso, H. Wunderlich, K. Modi, and A. Brodutch, as well as F. Buscemi for pointing out a mathematically equivalent statement of Thm.~(\ref{zerodiscord}) in~\cite{Hayashi06}, Lemma 8.12. Special thanks to C. M. Caves for directing my attention to an error in the previous version. This work was supported by the EPSRC grant EP/C546237/1 and the EU Integrated Projects QAP and QESSENCE. This work was completed while the author was at Imperial College, London.

\vspace{-0.0cm}

\bibliography{JCdiscord}

\end{document}